\documentclass[prc,twocolumn,english,showpacs]{revtex4}
\usepackage[T1]{fontenc}
\usepackage[latin1]{inputenc}
\usepackage{babel}
\usepackage{graphics}
\usepackage{comment}

\makeatletter

\providecommand{\LyX}{L\kern-.1667em\lower.25em\hbox{Y}\kern-.125emX\@}
\let\SF@@footnote\footnote
\def\footnote{\ifx\protect\@typeset@protect
    \expandafter\SF@@footnote
  \else
    \expandafter\SF@gobble@opt
  \fi
}
\expandafter\def\csname SF@gobble@opt \endcsname{\@ifnextchar[
  \SF@gobble@twobracket
  \@gobble
}
\edef\SF@gobble@opt{\noexpand\protect
  \expandafter\noexpand\csname SF@gobble@opt \endcsname}
\def\SF@gobble@twobracket[#1]#2{}

\usepackage[T1]{fontenc}
\usepackage[latin1]{inputenc}
\usepackage{babel}
\usepackage{graphics}



\makeatother
\begin{document}

\title{\( ^{8} \)B breakup in elastic and transfer reactions}
\author{A.M. Moro}
\email{moro@romantico.us.es}
\author{R. Crespo}
\affiliation{Departamento de F\'{\i}sica, Instituto Superior T\'{e}cnico,
Av. Rovisco Pais, 1049-001, Lisboa, Portugal}
\author{F. Nunes}
\affiliation{Universidade Fernando Pessoa, Pra\c{c}a 9 de Abril, 4200,
Porto, Portugal}
\author{I.J. Thompson}
\affiliation{Departament of Physics, University of Surrey, Guildford GU2 5XH, UK}

\begin{abstract}
We have studied the transfer reaction $^{14}$N($^7$Be,$^8$B)$^{13}$C 
at \( \mathrm{E}_{\mathrm{lab}}=84 \)~MeV, paying  special attention
to the effects of the coupling to the continuum in the exit channel.
Using the Continuum Discretized Coupled Channels (CDCC) formalism, we
find that these effects are important for the description of the elastic
scattering observables. However, for the transfer
process, differences between
the predictions of the differential cross section within
the Distorted Wave Born Approximation (DWBA)
and the CDCC-Born approximation (CDCC-BA)
are found to be negligible. This result supports the use
of the DWBA method as a reliable tool to extract the
$S_{17}(0)$ factor in this case.
\end{abstract}
\maketitle

\section{Introduction}

Many reaction formalisms have been developed to study stable and unstable
nuclei in the last few years and applied to extract structure information
from scattering processes. Often, there is an interplay
between the structure information that should be extracted from the
reaction data and the reaction process itself. Thus, in each case,
in order to extract reliable structure information, the adequacy of
the scattering formalism needs to be addressed in detail.

Of timely importance is the coupling to breakup states when the scattering
process involves loosely bound nuclei. Early analysis of elastic scattering
with deuterons have shown that it is important to include the couplings
to the continuum \cite{Ron70,Wat58}.
More recently, the analysis of scattering reactions of loosely bound nuclei
has progressed beyond the collective optical model (OM) approach
to a more microscopic treatment (i.e. \cite{Ron97b}
for elastic, \cite{Cres01}
inelastic, \cite{Ber95,Tos98a,Nun99,Tos01}
breakup and \cite{Tim99}
transfer reactions). In these approaches,
the few body nature of the loosely
bound nucleus is incorporated \emph{ab initio} in the scattering model
and the coupling to the breakup channels are either introduced explicitly
\cite{Yah82,Yah86,Aus87}, effectively through polarization potentials
\cite{May94,May95} or to all orders \cite{Cres99a,Mor01a}.

The study of the coupling to the continuum in transfer processes with
loosely bound nuclei is also of relevance for astrophysics. The asymptotic
normalization coefficient (ANC) method \cite{Xu94} has been put forward
as an alternative way to obtain information about the low energy \( S \)-factors.
This method uses the absolute normalization of a peripheral transfer
reaction to determine the normalization of the vertices involved in
the process. Its applicability depends crucially on the validity of
the DWBA conventionally used. The main assumptions are that the transition
amplitude for the transfer can be evaluated in Born approximation
and that the incoming and outgoing elastic waves are properly described
in terms of effective optical potentials. Typically, the optical potentials
used in DWBA calculations are deduced from the analysis of entrance
(and exit, whenever existent) elastic scattering data, to reduce the
uncertainties of the ANCs extracted from the transfer \cite{Fer99}.
The possibility of systematic errors in this method has been a source
of concern \cite{Alder98}. As a result, several tests have been performed
to ensure its validity (see for example a comparison with direct measurements
\cite{Gag99} or the extraction of the same information from a set
of different reactions \cite{Fer00}). Very recently, the importance
of coupling to excited inelastic channels of the target was assessed
\cite{Nun01} and results emphasize that care should be taken when
choosing the target.

A significant number of the unmeasured capture reactions of interest
in astrophysics involve nuclei on the drip line \cite{cosmos98,Alder98}.
The proximity of threshold suggests
that breakup channels may play a role in their reaction mechanism. These
implications have not yet been evaluated for any of the transfer reactions
used so far. A prime example is the extraction of \( S_{17} \) from
$^{14}$N($^{7}$Be,$^{8}$B)$^{13}$C
at \( \mathrm{E}_{\mathrm{lab}} \)=84~MeV \cite{Azh99,Trac00}.

One of the reasons for $^{8}$B attracting much of the nuclear
physics efforts is its relevance to astrophysics, namely to the solar
neutrino problem \cite{Alder98}. The first experiment performed with
the aim of extracting an ANC for $^8$B was a $(d,n)$ 
reaction \cite{Liu96}.
Meanwhile two transfer reactions on medium mass targets
were measured  with the same aim:
\( ^{10}\mathrm{B}(^{7}\mathrm{Be},^{8}\mathrm{B})^{9}\mathrm{Be} \)
\cite{Azh99,Azh99b} and 
$^{14}$N($^{7}$Be,$^{8}$B)$^{13}$C
\cite{Azh99} both at \( \mathrm{E}_{\mathrm{lab}} \)=84~MeV. The
joint analysis of these reactions \cite{Azh01} provided an accuracy
for \( S_{17}(0) \) greater than that of the direct capture 
measurements \cite{Ham01}.
Coupled channel estimates \cite{Nun01} showed that the excited states
of the target can have a strong influence when the target is \( ^{10} \)B
but not when $^{14}$N is used. Consequently, the \( S_{17}(0) \)
value extracted from 
$^{10}$B($^{7}$Be,$^{8}$B)$^{9}$Be
should not be used without further inelastic studies, but the value
extracted from 
$^{14}$N($^{7}$Be,$^{8}$B)$^{13}$C
remains valid up to now. Until recently, \( S_{17} \) extracted from
both the transfer reaction of \cite{Azh99} and the Coulomb Dissociation
data \cite{Dav01} were consistent with the direct capture measurements.
However, the very recent direct capture data from Seattle \cite{Jun02} not
only improves the accuracy, but provides a \( S_{17}(0) \) \( 30\% \)
larger than the previous values. Whilst differences within the direct
capture data sets are being understood, all sources of possible systematic
errors in the analysis of the data from indirect methods need to be
checked. Given that, in \( ^{8} \)B, the proton is bound by only
0.137 MeV, one can suspect that coupling to continuum states may
play an important role in the reaction dynamics and affect the ANC
results.

The aim of this work is thus to study consistently 
the effects of continuum couplings
of $^{8}$B, both in the elastic scattering
and the transfer process. In section II we discuss the formalism used
in both kinds of processes. In section III we analyze the elastic
scattering of $^{8}$B+$^{13}$C and discuss the
effects of coupling to the continuum in the calculated differential
cross section. In section IV, we analyze the transfer reaction
$^{14}$N($^{7}$Be,$^{8}$B)$^{13}$C
using the CDCC-BA framework: Born approximation (BA) for the transfer
couplings and Coupled Channel Continuum Discretization (CDCC) for
the \( ^{8} \)B continuum couplings. Finally in section V the conclusions
of the work are drawn.

\section{Theoretical framework} \label{sec:cdcc}

As mentioned previously, \( ^{8}\mathrm{B} \) is a weakly bound system
with a breakup $p$+$^{7}$Be threshold close to the ground
state. For many purposes this nucleus can be well described by a two
cluster model in which the valence proton is 
coupled to a $^{7}$Be
inert core (e.g. \cite{Esb96}). In order to 
include the $^{8}$B
continuum states as intermediate steps in the elastic or transfer
process we make use of the CDCC formalism \cite{Yah82,Yah86,Aus87}.

Consider the 3-body scattering problem of a composite nucleus \( A \)
= \( {\mathcal{C}}+v \) impinging on a stable nucleus \( c \). The
full Hamiltonian for the problem is \begin{equation}
H=h_{c}+h_{A}+T_{\alpha }+V_{\alpha }\; ,
\end{equation}
 where the Hamiltonian for the composite 
nucleus \( A \) is \( h_{A}=T_{{\mathcal{C}}v}+V_{{\mathcal{C}}v}+h_{\mathcal{C}}+h_{v} \)
and \( V_{\alpha } \) is the sum of interaction between the clusters
and the stable nucleus \( c \), \( \; V_{\alpha }=V_{c{\mathcal{C}}}+V_{vc} \).
Let $r$ be the $vc$ separation, and $R$ the projectile-target
coordinate.

For simplicity, we ignore the internal spins in the notation, and
consider only excitations of the projectile \( A \). Keeping in mind
the application to loosely bound nuclei, we assume that \( A \) has
only one bound state. Then \( h_{A}\phi _{0}(r)=\epsilon _{0}\phi _{0}(r) \)
defines the ground state wavefunction and 
\( h_{A}\phi _{\ell m,k}=\epsilon _{k}\phi _{\ell m,k} \)
defines the continuum states (labeled by the angular momentum \( \ell  \),
its projection \( m \), and the linear momentum, \( k \)).

The CDCC method makes a double truncation of the continuum in both
energy and angular momentum, working in the subspace \( 0\leq k\leq k_{max} \)
and \( 0\leq \ell \leq \ell _{max} \). Moreover, the excitation energy
range is subdivided into a number of intervals, usually called \emph{bins}.
For each such bin a representative square integrable wavefunction
is constructed, by an appropriate superposition of the continuum functions
inside the bin. Thus, for a total angular momentum \( J \) with projection
\( M \), the CDCC scattering wave function for the \( c+A \) system
is expanded as
\begin{eqnarray}
\label{eq:wfcdcc}
\Psi ^{CDCC}_{JM}(\mathbf{r},\mathbf{R})  =
[\phi _{0}(\mathbf{r})\otimes Y_{L}(\hat{R})]_{JM}\chi ^{J}_{0,L}(R)
\nonumber \\
+
\sum ^{\ell _{max}}_{\ell =0}\sum ^{J+\ell }_{L=|J-\ell |}\sum ^{N}_{i=1}\chi ^{J}_{i,L}(R)\left[ \phi _{i,\ell }(\mathbf{r})\otimes Y_{L}(\hat{R})\right] _{JM}\, ,
\end{eqnarray}
 with \( N=k_{max}/\Delta k \). Here \( \phi _{i,\ell }(\mathbf{r}) \)
are the bin wavefunctions; \( \chi ^{J}_{i,\ell L}(R) \) and
\( \chi _{0,L}^{J}(R) \) are the radial wavefunctions for the relative
motion between \( c \) and $A$.

The radial functions \( \chi ^{J}_{0,L} \) and \( \chi ^{J}_{i,L} \)
are determined by solving a set of coupled equations in the truncated
space. The coupling potentials between different channels are given
by:\begin{equation}
\label{eq:Vcoupl}
V_{i\ell :i'\ell '}(\mathbf{R})=\langle \phi _{i\ell }(\mathbf{r})|V_{\alpha }(\mathbf{r},\mathbf{R})|\phi _{i'\ell '}(\mathbf{r})\rangle ,
\end{equation}
 where it is understood that \( i \)=0 stands for the ground state.
These coupling potentials include the g.s-g.s matrix element (also
known as the Watanabe potential), g.s.-continuum terms and continuum-continuum
couplings. The latter can be handled in the same way as the
others
because the continuum bins have been made square integrable. The wavefunction
(\ref{eq:wfcdcc}) permits the description of the elastic and breakup
processes for the reaction \( A+c \).

Let us now consider the transfer reaction \( a(\mathcal{C},A)c\, \, \, \, (a=c+v,\, A={\mathcal{C}}+v) \),
where we have chosen the notation such that the exit channel of the
transfer corresponds to the elastic channel presented above. The prior
form of the transition amplitude for the transfer reaction process
is \cite{Sat83}: \begin{equation}
\label{Eq:Tprior}
T_{\textrm{prior}}=
\langle \Psi _{f}^{(-)}|V_{vc}+U_{c\mathcal{C}}-
U_{\alpha }|\chi _{\alpha }^{(+)}\phi _{a}\phi _{{\mathcal{C}}}\rangle ,
\end{equation}
 where \( V_{vc} \) is the the potential which binds the \( v \)
valence particle to the \( c \) core, \( U_{c\mathcal{C}} \) is
the core-core potential, and \( U_{\alpha } \) is an arbitrary potential
that generates the distorted wave function in the
entrance channel \( \chi _{\alpha}^{(-)} \).
Note that \( \Psi _{f}^{(-)} \) is the total exact wave function
with outgoing boundary conditions. In Eq.~(\ref{Eq:Tprior}), \( V_{vc} \)
is a real potential, while \( U_{\alpha } \) can be chosen either
as real or complex. It is usually chosen as the optical potential
\(U_{\alpha }= U_{a\mathcal{C}} \)  which reproduces the elastic
scattering in the entrance channel.

In practice, Eq.~(\ref{Eq:Tprior}) is not directly used, as it requires
the knowledge of the exact solution of the 3-body Schr\"{o}dinger
scattering equation for the entrance partition, a rather complicated
problem on its own. Approximations of this general form are developed
according to the desired applications.

When the coupling between the partitions \( a+\mathcal{C} \) and
\( A+c \) is sufficiently weak, the transfer process can be treated
in Born approximation. Even in this situation, if the
coupling to some excited states of any
of the partitions is strong, the transfer process can proceed via
these intermediate states. In these circumstances, it is convenient to
solve the coupled equations that include the couplings between the
different excited states, followed by the calculation of the transfer
in Born approximation. This procedure is known as the CCBA method
(Coupled Channels Born Approximation). When the coupling between
the excited states of the same partition are weak,  a further
approximation is commonly performed, by neglecting the explicit coupling
to these states.
This procedure is known as Distorted
wave Born approximation (DWBA). Even when the couplings are not so
weak, they may be at least partially taken into account by an appropriate
choice of the optical potentials.

When weakly bound nuclei are involved,  the CCBA is expected
to be a reasonable approach, because the transfer cross section
is small due to the
unfavorable $Q$--matching. However, couplings
to continuum states may still be important,
as these can act as intermediate steps in the
rearrangement process. It is not obvious that the
DWBA approach properly accounts for these continuum effects.
Nevertheless, this procedure has been  used recently in the analysis
of reactions involving weakly bound systems as a tool to extract
the ANC information. Therefore, it is timely to perform
a detailed analysis of the scattering frameworks used to described
the transfer processes. In particular, it is relevant to investigate
to what extent the effects of the coupling to the continuum can be
incorporated effectively in the optical potentials used by the DWBA approach.


In the case of elastic and inelastic scattering,
a common procedure is to represent the continuum spectrum
in \( \Psi _{f}^{(-)} \) by a finite
set of normalizable states, such as the CDCC expansion of
Eq.~(\ref{eq:wfcdcc}).
When rearrangement
channels are to be considered, as in Eq.~({\ref{Eq:Tprior}),
because the transfer step itself is not strong we can plausibly approximate the
exact wavefunction appearing in Eq.~(\ref{Eq:Tprior}) by the CDCC
wavefunction of Eq.~(\ref{eq:wfcdcc}). This procedure is called the
CDCC-BA method. In CDCC-BA the exact transition amplitude is approximated
by: 
\begin{equation}
T^{CDCC-BA}_{\textrm{prior}} =
 \langle \Psi _{f}^{\rm CDCC(-)}|V_{vc}+U_{c\mathcal{C}}-
U_{a{\mathcal{C}}}|\chi _{\alpha }^{(+)}\phi _{a}\phi _{{\mathcal{C}}}\rangle \; .\label{Eq:CDCC-BA}
\end{equation}
All couplings included in the CDCC-BA applied to
the (\( ^{7} \)Be,\( ^{8} \)B) case are schematically illustrated
in Fig.~\ref{Fig:b8_coup}.
\begin{figure}
{\centering \resizebox*{0.48\textwidth}{!}{\includegraphics{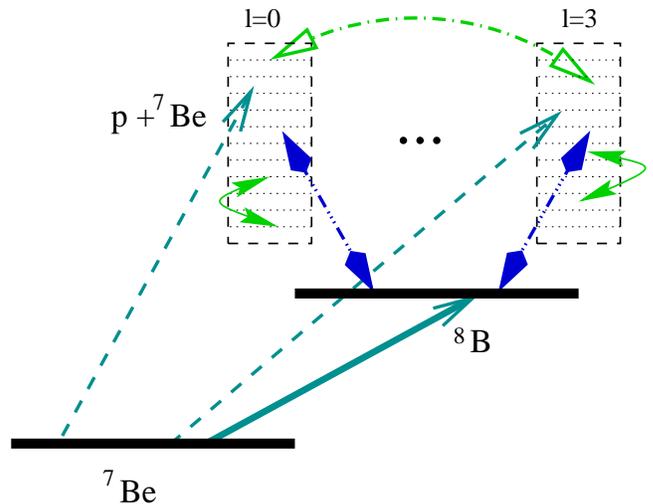}} \par}
\caption{\label{Fig:b8_coup}Couplings included in the CDCC-BA calculation.}
\end{figure}

Our choice of the prior representation for the transfer matrix element
is not arbitrary. In post form, the binding interaction is that of the
exiting projectile \( V_{v\mathcal{C}} \) and the remnant term is
the difference between the core-core interaction \( U_{c\mathcal{C}} \), and
the optical potential for the exit channel $U_{a{\cal{C}}}$.
However, this exit channel optical potential is often
unknown: contrary to the entrance channel elastic data, elastic
scattering in
the exit channel cannot be measured in the same experiment, and in many cases
it is just  unmeasurable as both projectile and target are radioactive.
It can then be
found either by fitting the scattering from the CDCC model,
or by extrapolating from neighboring nuclei.
The prior form of the CDCC-BA approach, by contrast,
does not require any knowledge of the optical potential  \( U_{\beta } \)
for this exit channel.
It seems more appropriate, therefore, to use the prior form representation of
the transition operator and, consequently, all the calculations
presented hereafter were performed in this representation.

The further approximation of the CDCC by just its elastic
channel, found with some optical  potential  \( U_{\beta } \), 
gives the DWBA transition amplitude in prior form:
\begin{equation}
T^{DWBA}_{\textrm{prior}}= \langle
\chi _{\beta }^{(-)}\phi _{^{_{A}}}\phi _{c}|
V_{vc}+U_{c{\mathcal{C}}}-U_{a{\mathcal{C}}}|
\chi _{\alpha }^{(+)}\phi _{a}\phi _{{\mathcal{C}}}\rangle \; .\label{Eq:DWBA}
\end{equation}
This simple approximation, very commonly used in ANC
analyses, still requires knowledge of the optical potential for the exit
channel $U_{a{\cal{C}}}$, and hence suffers from the
difficulties enumerated above. However, if the transfer cross
sections are not very sensitive to this potential, then the
DWBA will still be a useful procedure. We examine this
sensitivity below, by 
studying the degree of agreement between the 
DWBA and the CDCC.
In both approaches,  the remnant term of the transition matrix element for
systems where a nucleon is transferred from a 
well bound state to a loosely bound state
(or vice-versa) is often not negligible and should be accounted for properly.

\begin{table}

\caption{\label{Table:OP}Parameters of the potentials used in this
work. Depths are expressed in MeV and radii and diffuseness in fm.
The first rows correspond to optical potentials and the bottom
rows are the binding potentials. Reduced radii
are to be multiplied by
\protect\protect\( A^{1/3}_{P}+A^{1/3}_{T}\protect \protect \)
for nucleus-nucleus and by \protect\protect\( A^{1/3}_{T}\protect \protect \)
for nucleon-nucleus scattering. }

\begin{tabular}{|c|cccccccccc|}
\hline System &
 \( V \)&
\( V_{so} \) &
 \( r_{V} \)&
 \( a_{V} \)&
 \( W_{V} \)&
\( W_{S} \)&
 \( r_{W} \)&
 \( a_{W} \)&
 \( r_{C} \)&
 Ref.\\
\hline
$^7$Be+$^{13}$C
 (1)  &
 54.3 &
&
 0.92 &
 0.79 &
 29.9 &
&
 1.03 &
 0.69 &
 1.&
 \cite{Trac00}\\
$^7$Be+$^{13}$C
 (2)  &
 99.8&
&
 0.77&
 0.81&
 22.0&
&
 1.01&
 0.81&
 1.&
 \cite{Trac00}\\
$p$+$^{13}$C&

 60.4&
&
 1.14&
 0.57&
&
 5.8&
 1.14&
 0.50&
 1.25&
 \cite{Wat69}\\
\hline
$p$+$^{7}$Be&

 44.7&
4.9 &
 1.25&
 0.52&
&
&
&
&
 1.25&
 \cite{Esb96}\\
 $p$+$^{13}$C&

 51.4&
&
 1.30&
 0.65&
 &
 &
 &
&
 1.30&

\\
\hline
\end{tabular}
\end{table}

\section{The elastic scattering $^{8}$B + $^{13}$C}

We investigate in this section the elastic scattering \( ^{8} \)B
+ $^{13}$C at E\( _{_{\mathrm{lab}}} \)=78.4 MeV, which is the
exit channel in the transfer reaction we wish to analyze in the present
work. In particular, we study the importance of the continuum
in the calculated
differential cross section.
This reaction was previously analyzed
in the work of \cite{Trac00} using a renormalized
double folding (RDF) potential obtained by an analysis of nearby stable
nuclei. A parameterization of this RDF potential in terms of usual
Woods-Saxon forms was also derived by fitting the outer part of the
RDF potential. Since these fitted potentials may differ in the inner
part from the original RDF potentials, we used the latter in our calculations,
as in the results presented in \cite{Trac00}.
As this interaction is derived from a systematic study on stable nuclei,
it is not clear how adequate are these extrapolations to loosely bound nuclei.
Elastic data for $^{8}$B + $^{13}$C would help to shed light
on these issues.

In this section we analyze the same reaction in terms of the CDCC
formalism, for two reasons. First,
this treatment allows an explicit study of the role of the continuum
and, second, it provides an alternative analysis
to the RDF which does not require the optical potential
for $^{8}$B + $^{13}$C, thus providing a valuable reference
in the absence of experimental data.

An important ingredient of the CDCC calculation is the bound state
wavefunction of the \( ^{8} \)B nucleus. For the binding potential,
we have adopted the parameters given in \cite{Esb96} and
listed in Table \ref{Table:OP}.
The valence proton wavefunction is considered to be a pure \( p_{3/2} \)
configuration coupled to a zero spin core of \( ^{7}\)Be with unit spectroscopic
factor. Although it is known that there is a
$p_{1/2}$ which has a small contribution to the cross section \cite{Azh99},
we chose to neglect it to make the CDCC calculations feasible.

The interaction between the projectile \( ^{8}\mathrm{B} \), and
the target $^{13}$C to be used in the CDCC calculation
is written as the sum of the interactions
$U$($^{7}$Be,$^{13}$C)
and $U$($p$,$^{13}$C). No experimental data for the elastic
scattering $^{7}$Be + $^{13}$C at the relevant
energies (\( \approx  \)68 MeV) have been found in the literature.
Nevertheless, the similarity in the structure of \( ^{7} \)Be and
its mirror partner \( ^{7} \)Li suggests to describe this reaction
using the potential taken from the
reaction $^{7}$Li+$^{13}$C},
for which experimental data exists at 63 MeV.
The $U$($p$,$^{13}$C)
was taken from nucleon-nucleus global parameterizations.
\begin{figure}
\resizebox*{0.48\textwidth}{!}{\includegraphics{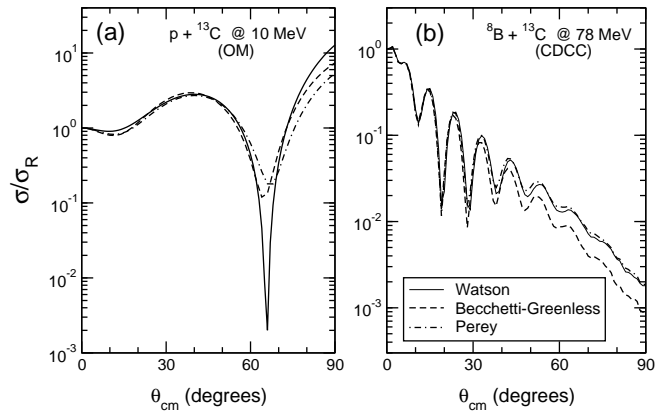}}
\caption{\label{Fig:b8c13_cdcc_fold}
a) $p$ + $ ^{13}$C
differential elastic cross section for different optical potentials.
b) CDCC differential elastic cross section angular distribution for
$^{8}$B+$^{13}$C
at 78 MeV using different parameterizations for 
the $p$ + $^{13}$C interaction.}
\end{figure}

Convergence of the CDCC results was achieved with matching radius of 40 fm
and maximum total angular momentum of \( L_{\mathrm{max}} \)=100.
The continuum spectrum was divided into \( N=10 \) bins of
equal energy width
in the range from 0 to 9 MeV. We took into
account \( s \), \( p \) and \( d \) continuum partial waves. All
the calculations were performed with the computer code FRESCO \cite{Thom88}.

We checked the sensitivity of the calculation with respect to the
$U$($^{7}$Be,$^{13}$C) and $U$($p$,$^{13}$C)
interactions. To analyze the uncertainty associated with the interaction
$U$($p$,$^{13}$C), we compare in Fig.~\ref{Fig:b8c13_cdcc_fold}(a)
the calculated differential cross section for the $p$+$^{13}$C 
elastic scattering (as ratio to Rutherford), 
using several proton-$^{13}$C
interactions, adopted from the global parameterizations of Watson
\cite{Wat69}, represented by the solid curve,
Becchetti-Greenless \cite{Bec69} (dashed line) and
Perey \cite{Per63} (dashed dotted line). In all three cases
the spin-orbit term was omitted.

The calculated CDCC elastic angular  distributions shown    
in Fig.~\ref{Fig:b8c13_cdcc_fold}(b)  
with the Watson and Perey parameterizations
are very similar, but significantly different from that calculated
with the Becchetti-Greenless potential.
As is well known,
the Becchetti-Greenless parameterization is
better suited for medium and heavy mass nuclei, and for higher scattering
energies. Thus, the use of this optical potential to describe the
scattering of \( p+^{13} \)C around 10 MeV may be questionable. This,
together with the fact that the Watson and Perey parameterizations
provide essentially the same elastic scattering,
suggests that we can use either of these with
some confidence. From hereafter, the Watson potential of
\cite{Wat69} will be used in all the CDCC calculations.
\begin{figure}
\resizebox*{0.48\textwidth}{!}{\includegraphics{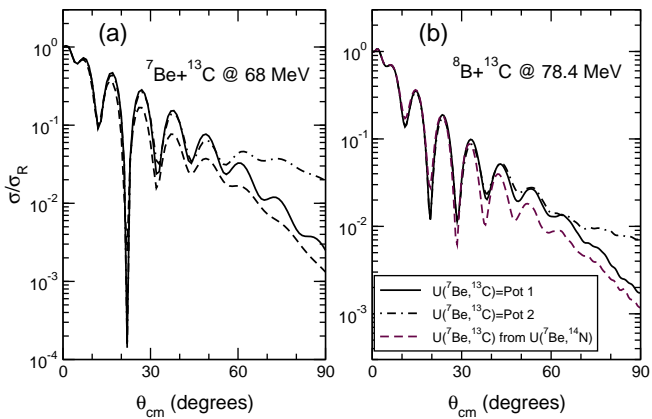}}
\caption{\label{Figure: cdcc_core} a) Calculated differential cross section
angular distribution for 
$^{7}$Be + $^{13}$C
at 68.6 MeV within the optical model formalism using different potential
parameterizations. b) Corresponding CDCC elastic scattering 
for $^{8}$B+$^{13}$C
at 78.4 MeV.}
\end{figure}

We study now the sensitivity of the calculated differential cross
sections with respect to the $U$($^{7}$Be,$^{13}$C)
potential. Fig.~\ref{Figure: cdcc_core}(a) shows
the calculated elastic scattering cross section
for $^{7}$Be+$^{13}$C using the potentials
Pot1 (solid curve), Pot2 (dashed-dotted curve) from the reaction
$^{7}$Li+$^{13}$C at 63 MeV analysed in \cite{Trac00}
and listed in Table \ref{Table:OP}. Also included in
Fig.~\ref{Figure: cdcc_core}(a)
is the result obtained with the optical potential
$U$($^{7}$Be,$^{14}$N) (dashed line).
In Fig.~\ref{Figure: cdcc_core}(b) we
show the corresponding CDCC calculations for the
$^{8}$B + $^{13}$C elastic scattering at 78.4 MeV.
It is observed that the
CDCC calculations for $^8$B$+^{13}$C, using potentials Pot1 and Pot2,
give very similar results, whereas when the core-target potential is
taken to be $U$($^{7}$Be,$^{14}$N),
a somewhat bigger difference is encountered beyond 30 degrees.
In the following, we will use Pot1 as the core-target interaction.

\begin{figure}
\resizebox*{0.45\textwidth}{!}{\includegraphics{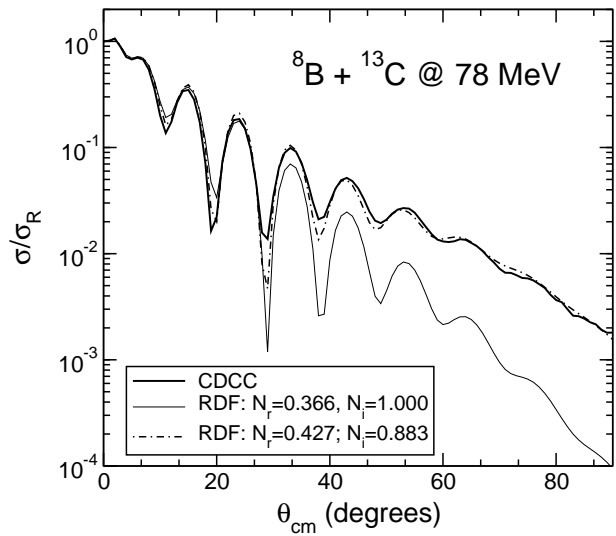}}
\caption{\label{Fig:CDCC_DF} Elastic cross section angular distribution (as
ratio to Rutherford cross section) calculated in the CDCC framework
(thick solid line) and renormalized double folding
with the real renormalization
constant proposed in Ref.~\cite{Trac00} (thin solid line)
and with renormalization
constant adjusted to fit the CDCC (dotted-dashed line). }
\end{figure}

Finally, we compare in Fig.~\ref{Fig:CDCC_DF} the
CDCC (thick solid line) and 
pure OM calculations. The thin solid line is the OM calculation, using
the RDF potential derived in \cite{Trac00}. This agrees very well
with the CDCC at small angles
(up to 25 degrees), but presents significant discrepancies
beyond this range.
Also included in this figure is an OM calculation using the same double
folding potential, but fitting the real and imaginary renormalization
constants to approximate
the CDCC result. We found that a
value of \( N_{r}=0.427\) and \(N_{i}=0.883 \)  (dotted-dashed line)
provides an excellent agreement between both calculations, in contrast
to the value \( N_{r}\simeq 0.366 \) (thin solid line)
proposed in \cite{Trac00}. Any experimental results
for this reaction at $\theta > 30$ degrees would 
help clarify the adequacy of global parameterizations
for loosely bound nuclei.

\begin{figure}
\resizebox*{0.48\textwidth}{!}{\includegraphics{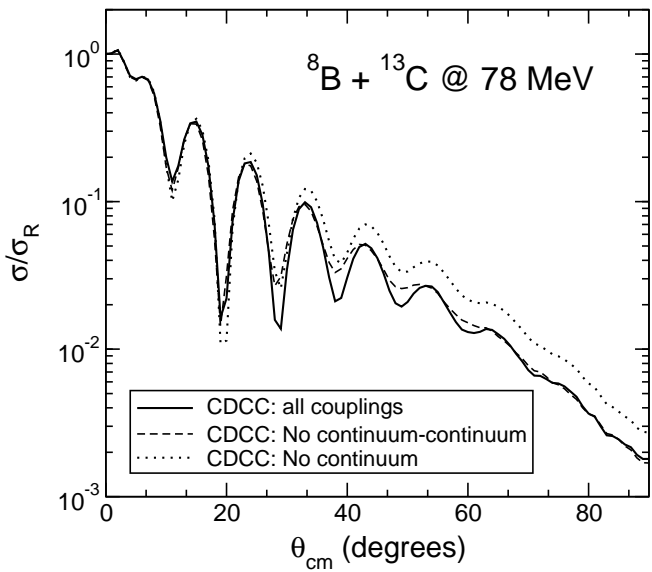}}
\caption{\label{Fig:b8c13_cont}Effect of the continuum in \protect\protect\( ^{8}\protect
\protect \)B~+~\protect\protect\( ^{13}\protect \protect \)C elastic scattering at 78
MeV. The dotted line is the Watanabe calculation (i.e., the CDCC calculation without
gs--continuum couplings). The dashed is the CDCC with gs--continuum couplings, but no
continuum--continuum couplings. The solid line is the full CDCC calculation with both
gs--continuum and continuum--continuum couplings. }
\end{figure}

We have also estimated the effect of the \( ^{8} \)B continuum
on the calculated
elastic scattering cross section. To this end we compare
in Fig.~\ref{Fig:b8c13_cont}
the elastic differential cross section using the full CDCC calculation
(thick solid line) with a calculation in which all couplings with
continuum states have been ignored (dotted line). The latter
is equivalent to a optical model
calculation in which the projectile--target interaction is described in
terms of the Watanabe folding potential. Also
represented is the CDCC without continuum--continuum couplings
(dashed line).
It can be seen that this truncated calculation provides cross sections
which are already very close to those produced by the
full CDCC calculation, suggesting that multistep processes
coupling different continuum states are not very relevant in this
reaction. The main CDCC effect appears to be a reduction of
the cross sections caused by absorption due to breakup at
near-grazing collisions.

\begin{figure}[t]
\resizebox*{0.48\textwidth}{!}{\includegraphics{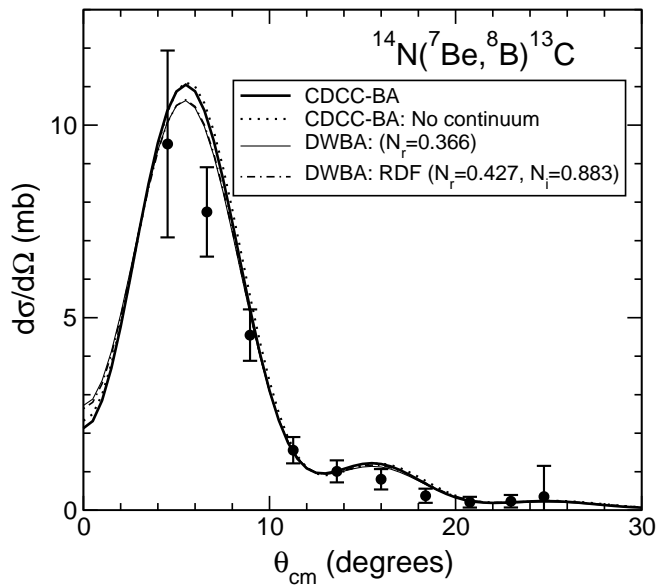}}
\caption{\label{Fig:DWBA-CDCC-BA}
Calculated transfer cross section angular distribution
for the reaction $^{14}$N($^{7}$Be,$^{8}$B)$^{13}$C at
84 MeV. The thick solid line corresponds to the full CDCC-BA calculation. The dotted line
is the CDCC-BA calculation without gs-continuum couplings. The thin solid line is the
DWBA calculation using the RDF optical potential for exit channel 
with \( N_{r}=0.366 \).
The dotted--dashed line is the DWBA calculation with the same RDF potential, but with the
complex renormalization constant \( N_{r}=0.427 \) and \(N_{i}=0.833.\) 
All the CDCC-BA calculations include the factor 5/16 
to account for the physical nuclei spins, as discussed in section III.
}
\end{figure}

\section{The transfer reaction $^{14}$N($^{7}$B\lowercase{e},$^{8}$B)$^{13}$C}

We analyze in this section the transfer reaction
$^{14}$N($^{7}$B\lowercase{e},$^{8}$B)$^{13}$C at 84 MeV, within
the DWBA and CDCC-BA approaches.
In order to make a reliable comparison between both formalisms we
have calculated the transition amplitudes in
Eq.~(\ref{Eq:DWBA}) and Eq.~(\ref{Eq:CDCC-BA})
using the same core-core interaction, \( U_{c\mathcal{C}} \):
we take  Pot1 from Table \ref{Table:OP}.
For the binding potential ($V_{vc}$) of $p$ + $^{13}$C
we used the parameters
listed in Table  \ref{Table:OP}.  
The entrance channel was described in terms
of the numerical RDF potential derived in \cite{Trac00}. 
The DWBA calculation requires also the
exit optical
potential $U_{\beta}$= $U$($^{8}$B,$^{13}$C),
which we also took from  \cite{Trac00} in numerical form.

The ground state and continuum structure of $^8$B needed for the
CDCC calculation
was taken to be the same as the one in the previous section.
The \( ^{14} \)N   ground state was described
as a proton in a \( p_{1/2} \) configuration, with spectroscopic
factor 0.604 \cite{Azh99}.
For the purpose of comparing the present calculations with the data
we have renormalized $all$ the calculated transfer cross sections
by the spectroscopic factor $S_{p_{3/2}}$=0.737. 
This value was derived from the ANC reported in
\cite{Azh99} for the \( p_{3/2} \) configuration
in the $^{8}$B ground state ($C^2_{p_{3/2}}$=0.371 fm$^{-1}$) 
and the calculated asymptotic
normalization constant for the single-particle orbital. 
All $^8$B continuum couplings are taken into account,
but no transfer back-couplings are included,
as this would worsen the fit to the elastic scattering
in the entrance channel.

Note also that our simplified description of $^8$B in terms of a proton
coupled to a zero spin core 
provides a cross section which,
after multiplication of the factor
($2I_A+1)(2I_c+1)/(2I_a+1)(2I_{\cal{C}}+1)$,
is equivalent
\footnote{In the absence of spin dependent forces,
the transfer cross section
for $^{14}$N+$^{7}$Be($0^+$)$\rightarrow ^{13}$C+$^{8}$B($3/2^-$)
should equal the sum of the
cross sections
$\sum_I^{14}$N+$^{7}$Be(3/2$^-$)
$\rightarrow ^{13}$C + $^{8}$B($I$), for
$I=0^+,1^+,2^+,3^+$
that result from the coupling of $I(^7\mathrm{Be})$
to the $p_{3/2}$ proton single particle
state. The physical process measured is only
$^{14}$N+$^{7}$Be($3/2^-$)$\rightarrow ^{13}$C+$^{8}$B($2^+$).
According to detailed balance, one can conclude that the weight of
this reaction relative to the total sum is
($2I_A+1)(2I_c+1)/(2I_a+1)(2I_{\cal{C}}+1)$=5/16. This means that the
calculated
transfer cross section using a simplified description with zero
spin for $^7$Be needs to be multiplied by this factor.
}
to the cross section calculated with correct spins.

The resulting transfer cross sections are presented
in Fig.~\ref{Fig:DWBA-CDCC-BA}.
The thick solid line corresponds to the full CDCC-BA calculation.
The thin solid line represents the DWBA calculation
with a renormalized double folding (RDF) potential for the entrance
and exit channels, using $N_r$=0.366 for the two potentials. The
resulting angular distribution is very close to the CDCC calculation,
differing by only 5\% at the maximum of the distribution, which is 
unmeasurable within the present experimental accuracy.

We stress however, that, as shown in the previous section, OM and CDCC
give different predictions for the elastic scattering
in the exit channel at large angles.
Under the circumstances, we believe that the CDCC predictions for the
elastic $^8$B$+^{13}$C is more reliable (see discussion in section III).
Notwithstanding, we have shown that these coupling effects can be
easily included in the optical potential by adjusting 
the renormalization constants.
We have then performed a DWBA calculation using
the RDF with a complex renormalization constant
\( N_{r}=0.427, N_{i}=0.883 \),
accurately reproducing the CDCC elastic predictions
(dot--dashed line in Fig.~\ref{Fig:DWBA-CDCC-BA}).
Despite the fact that
this change in the renormalization constants modifies 
significantly the elastic cross sections of the exit channel
beyond 30 degrees, the resulting transfer cross section remains
basically unaltered up to 25 degrees,
the angular range used in 
\cite{Azh99} to extract the ANC information. 
This result seems to support the 
peripheral nature of this reaction.
Also shown in the figure is the calculated transfer cross section
without continuum couplings (dotted line). This appears to be very
similar to the full CDCC calculation. 
The similarity between all curves
demonstrates the minor role played by the continuum of
$^{8}$B and confirms the DWBA formalism as an adequate
tool to extract the ANC in this reaction, as done in \cite{Azh99}.

In order to check the dependence of this result with 
the bombarding energy, 
we compared the CDCC and DWBA 
at other energies. We found that the effect of the continuum 
decreases as the incident energy increases. For instance, at 160 MeV 
the transfer cross sections calculated in CDCC-BA 
and DWBA differ by less than 1\% at angles up to 30 degrees.
By contrast, at 40 MeV differences of around 8\% where found at the maximum
of the angular distribution.

\section{Conclusions }

In summary, we have studied the reaction 
$^{14}$N($^{7}$Be,$^{8}$B)$^{13}$C
at 84 MeV, placing special stress on the importance of the continuum
of \( ^{8} \)B in the description of the exit channel and in the
transfer process. This reaction has been recently measured and analyzed
within the DWBA formalism \cite{Azh99}, in order to extract the astrophysical
\( S_{17} \) factor for the capture reaction \( ^{7}\mathrm{Be}+p\rightarrow ^{8}\mathrm{B} \).
The validity of this procedure relies on the assumption that the transfer
reaction occurs in one step and, also, that the entrance and exit
channels are well described by optical potentials.

The importance of the \( ^{8} \)B continuum in the reaction mechanism
has been analyzed by describing the \( ^{8} \)B+\( ^{13} \)C scattering
in terms of \( p + ^{13} \)C and \( ^{7} \)Be+\( ^{13} \)C
optical potentials and discretizing the \( ^{8} \)B continuum into energy bins.
The elastic cross sections given by the CDCC solution has been compared
with those obtained in the optical model analysis performed in \cite{Trac00}.
In this reference, the elastic scattering of \( ^{8} \)B+\( ^{13} \)C was
analyzed in terms of a double folding optical potential,
using a renormalization
constant derived from a systematic analysis of several reactions involving
stable nuclei in the same energy and mass region. We found that the
calculated differential elastic cross section is very similar in both
approaches at forward angles, but they differ significantly at larger
angles. This result cast doubt on the extrapolation of the global optical
potentials derived from stable nuclei to loosely bound nuclei.
Interestingly, in this reaction the CDCC effects can be accounted for
very well by correcting the normalization of the double folding potential.
The calculated CDCC wave function is then used  in the expression
for the transfer amplitude.
Despite the discrepancies on the elastic scattering of the exit channel
at large angles,
the calculated transfer cross sections are very similar in the DWBA
and CDCC-BA approaches below 25 degrees, which was the angular range
used to extract the ANC for this reaction.
Taking into account the result of \cite{Nun01},
where coupling to excited (bound) states was also found to be negligible,
to the accuracy of the present transfer data,
the present analysis supports the validity of the DWBA method as a
reliable tool to extract the \( S \)--factor from the
$^{14}$N($^{7}$Be,$^{8}$B)$^{13}$C reaction at the studied energy.
Consequently, these higher order corrections to the DWBA cannot justify the
disagreement between the $S_{17}$ extracting using the ANC method and
the new direct results from Seattle \cite{Jun02}.

Similar checks for other reactions that involve loosely bound nuclei
are underway. As the structure of $^8$B was simplified, some 
interference effects
could not be probed in this work. Even if less important when continuum
coupling effects are small, these effects should generally be properly
included. This problem is also being addressed.

\acknowledgments{We are deeply grateful to
J. G\'{o}mez-Camacho for his fruitful comments, and to
L. Trache for providing the data, the numerical optical potentials
and other details concerning the analysis of the data presented.
Support from Funda\c{c}ao para a Ci\^encia e a Tecnologia (F.C.T.) 
under the grant
SAPIENS/36282/99 and EPSRC under grant GR/M/82141 are acknowledged.
One of the authors (A.M.M.) acknowledges a F.C.T. post-doctoral grant.}


\end{document}